\documentclass{emulateapj}
\usepackage{natbib}
\usepackage{xspace}

\newcommand{\water}{H$_2$O\xspace}
\newcommand{\kms}{ km\,s$^{-1}$\xspace}

\newcommand{\miriad}{{\sc Miriad}\xspace}

\shorttitle{Unusual shock-excited OH masers in a young PN}
\shortauthors{Qiao et al.}

\begin{document}

\title{Unusual shock-excited OH maser emission in a young planetary nebula}

\author{Hai-Hua Qiao\altaffilmark{1, 2, 3, 4}, Andrew J. Walsh\altaffilmark{3}, Jos\'e F. G\'omez\altaffilmark{5}, Hiroshi Imai\altaffilmark{6}, James A. Green\altaffilmark{7}, Joanne R. Dawson\altaffilmark{8, 9}, Zhi-Qiang Shen\altaffilmark{1, 4}, Simon P. Ellingsen\altaffilmark{10}, Shari L. Breen\altaffilmark{9}, Paul A. Jones\altaffilmark{11}, Steven J. Gibson\altaffilmark{12}, Maria R. Cunningham\altaffilmark{11}}

\altaffiltext{1}{Shanghai Astronomical Observatory, Chinese Academy of Sciences, 80 Nandan Road, Shanghai, China, 200030; haihua.qiao@curtin.edu.au} 
\altaffiltext{2}{University of Chinese Academy of Sciences, 19A Yuquanlu, Beijing, China, 100049} 
\altaffiltext{3}{International Centre for Radio Astronomy Research, Curtin University, GPO Box U1987, Perth WA 6845, Australia}
\altaffiltext{4}{Key Laboratory of Radio Astronomy, Chinese Academy of Sciences, China}
\altaffiltext{5}{Instituto de Astrof\'{\i}sica de Andaluc\'{\i}a, CSIC, Glorieta de la Astronom\'{\i}a s/n, E-18008, Granada, Spain} 
\altaffiltext{6}{Department of Physics and Astronomy, Graduate School of Science and Engineering, Kagoshima University, 1-21-35 Korimoto, Kagoshima 890-0065, Japan}
\altaffiltext{7}{SKA Organisation, Jodrell Bank Observatory, Lower Withington, Macclesfield, Cheshire SK11 9DL, UK}
\altaffiltext{8}{Department of Physics and Astronomy and MQ Research Centre in Astronomy, Astrophysics and Astrophotonics, Macquarie University, NSW 2109, Australia}
\altaffiltext{9}{Australia Telescope National Facility, CSIRO Astronomy and Space Science, P.O. Box 76, Epping, NSW 1710, Australia}
\altaffiltext{10}{School of Physical Sciences, Private Bag 37, University of Tasmania, Hobart 7001, TAS, Australia}
\altaffiltext{11}{School of Physics, University of New South Wales, Sydney, NSW 2052, Australia}
\altaffiltext{12}{Department of Physics and Astronomy, Western Kentucky University, 1906 College Heights Blvd, Bowling Green, KY 42101 U.S.A.}
\begin{abstract}

We report on OH maser emission toward G336.644$-$0.695 (IRAS 16333$-$4807), which is a \water maser-emitting Planetary Nebula (PN). We have detected 1612, 1667 and 1720\,MHz OH masers at two epochs using the Australia Telescope Compact Array (ATCA), hereby confirming it as the seventh known case of an OH-maser-emitting PN. This is only the second known PN showing 1720\,MHz OH masers after K 3$-$35 and the only evolved stellar object with 1720\,MHz OH masers as the strongest transition. This PN is one of a group of very young PNe. The 1612\,MHz and 1667\,MHz masers are at a similar velocity to the 22\,GHz \water masers, whereas the 1720\,MHz masers show a variable spectrum, with several components spread over a higher velocity range (up to 36\kms). We also detect Zeeman splitting in the 1720\,MHz transition at two epochs (with field strengths of $\sim 2$ to $\sim 10$ mG), which suggests the OH emission at 1720\,MHz is formed in a magnetized environment. These 1720\,MHz OH masers may trace short-lived equatorial ejections during the formation of the PN.

\end{abstract}

\keywords{ masers -- planetary nebulae: general -- planetary nebulae: individual (IRAS 16333$-$4807) }

\section{Introduction}
\label{introduction}

OH maser emission arises from a wide range of astrophysical objects, such as high-mass star forming regions \citep[e.g.][]{Are2000}, evolved stars \citep[e.g.][]{Nge1979}, supernova remnants \citep[SNRs; e.g.][]{Fre1996}, comets \citep{Gee1998} and active galactic nuclei \citep{Bae1982}. In the particular case of evolved stars, ground-state OH maser emission at 18 cm is widespread in the circumstellar envelopes of oxygen-rich stars during the asymptotic giant branch (AGB) phase \citep{EB2015}, which is characterised by strong mass loss {\citep[up to $\simeq 10^{-4}$ M$_\odot$ yr$^{-1}$,][]{vas93,blo95}}. The existence of ground-state OH masers in evolved stars is indicative of highly energetic processes (e.g. strong radiation fields and shocks), which are required to invert the energy levels involved in these transitions.

Circumstellar envelopes of low and intermediate mass (0.8-8$M_{\odot}$) stars usually exhibit prominent OH maser emission at 1612\,MHz. The spectra of these masers during the AGB phase tend to show double-horned profiles, with two peaks indicating expansion velocities typically between 10 to 30\kms, tracing the expansion of the circumstellar envelope \citep[e.g.][]{See1997}. Also, some evolved stars show maser emission at 1665 and 1667\,MHz \citep[e.g.][]{Hue1994,Dee2004}, the velocities of which are close to that of the 1612\,MHz masers \citep{Dee2004}. The 1720\,MHz OH masers are best known to be associated with shocks created by SNRs interacting with surrounding molecular clouds \citep{Fre1996,Gre1997}. The 1720\,MHz maser transition has never been detected in AGB stars and is extremely rare in other evolved sources, i.e. it has only been detected in one post-AGB star \citep[OH009.1$-$0.4; IRAS 18043$-$2116;][]{Seb2001} and one Planetary Nebula (PN) \citep[K 3$-$35;][]{Goe2009}. \citet{Seb2001} argued that the 1720\,MHz masers from OH009.1$-$0.4 might arise in the shock between the AGB superwind and the fast post-AGB winds. \citet{Wae2009} also detected extremely high velocity \water masers towards this source and suggested they originated in a jet.

Stars enter the post-AGB phase when the AGB mass loss ceases. The OH maser emission from the expanding circumstellar envelope in the post-AGB phase is expected to disappear on a timescale of around 1000 years \citep{Lew1989,Goe1990}. The PN phase comes after the post-AGB phase, when the central star is hot enough to photoionize the surrounding envelope. In the PN phase, there are significant changes in the envelope kinematics and morphology compared to the post-AGB phase.
Compared to the copious OH masers in the AGB phase, the prevalence of OH masers decays significantly in subsequent phases, and they are rarely found in PNe. So far, only six PNe have been confirmed to harbour OH masers \citep[OH-maser-emiting PNe, hereafter referred to as OHPN/OHPNe;][]{Zie1989,Use2012}. They all show maser emission with strong asymmetry, i.e. preferentially blue-shifted emission with respect to the stellar velocity \citep{Use2012}. One of them, K 3$-$35, shows OH maser emission at all four ground-state transitions \citep[1612, 1665, 1667, 1720\,MHz;][]{Goe2009}, as well as excited-state masers at 6035\,MHz \citep{Des2010}. It is so far the only PN known to exhibit OH maser lines at 1720\,MHz. The velocity and location of the 1720\,MHz OH maser emission are also close to those of 22\,GHz \water masers, which indicates that these two kinds of masers may be excited by the same shock \citep{Goe2009}.

The Southern Parkes Large-Area Survey in Hydroxyl (SPLASH) is a survey of the OH radical across the inner Galaxy to high sensitivity \citep{Dae2014}. In order to determine accurate positions for OH masers detected in SPLASH, we have used the Australia Telescope Compact Array (ATCA), which allows us to obtain arcsecond positional accuracy of the masers. In this paper, we report on one OH maser source (G336.644$-$0.695; hereafter referred to as IRAS 16333$-$4807), seen at 1720\,MHz as well as 1612 and 1667\,MHz, that is found towards a previously-known PN, thus making it the second example of this very small class of 1720\,MHz OH maser sources after K 3$-$35.

\section{Observations and Data Reduction}
\label{observation}

We observed OH maser candidates detected in previous Parkes observations of the SPLASH pilot region (between Galactic longitudes of $334^{\circ}$ and $344^{\circ}$ and Galactic latitudes from $-2^{\circ}$ to $+2^{\circ}$) using the ATCA in October 2013 (6A configuration). The Compact Array Broadband Backend (CABB) was used to collect spectral line data, using the CFB 1M-0.5k mode with 16 `zoom' bands, each with 2048 channels over 1\,MHz giving 0.5 kHz channel spacings, with all four polarisation products. Each maser candidate region was observed for 30 minutes with five snapshot observations of duration 6 min each. The CABB system provides two IF bands of $\simeq$2\,GHz each, which for these observations were both centred at 2.1\,GHz. Further details of the observations will be published shortly, i.e. a catalogue of ground-state OH maser sources from the SPLASH pilot region (Qiao et al. 2015, {\em in preparation}). Radio-frequency interference (RFI) during the observations adversely affected the 1667\,MHz spectrum for IRAS 16333$-$4807 and because of this we made additional observations of the source in May 2015 (1.5C configuration). The setup was the same as the first epoch. We used PKS 0823$-$500 as the bandpass calibrator, PKS 1934$-$638 as the flux density calibrator and PKS 1613$-$586 as the phase calibrator.

The spectral line data (the `zoom' bands) were calibrated and imaged with the \miriad package using standard procedures \citep{SK2004} to obtain accurate positions and spectra. Dirty maps were obtained with the task \textsc{invert} of \miriad without down-weighting (robust=$-$infinity). The beamsizes are $7.9\arcsec\times5.0\arcsec$ (first epoch) and $13.4\arcsec\times5.8\arcsec$ (second epoch) for 1612\,MHz, $7.6\arcsec\times4.7\arcsec$ (first epoch) and $13.0\arcsec\times5.6\arcsec$ (second epoch) for 1667\,MHz and $7.4\arcsec\times4.5\arcsec$ (first epoch) and $12.6\arcsec\times5.4\arcsec$ (second epoch) for 1720\,MHz. Images were subsequently deconvolved with the tasks \textsc{clean} and \textsc{restor}. Correction for the primary beam response was also applied to the data using the task \textsc{linmos}. The task \textsc{imfit} was used to fit the integrated intensity map of each maser spot with a Gaussian. This provided both the position and relative positional uncertainty of the maser spots (which is smaller than the absolute positional uncertainty). The absolute positional uncertainty can be as large as $\simeq$10 per\,cent of the beam size for weak masers. We binned over three channels to smooth the line data for a final spectral resolution of $\sim$0.27\kms. A spatial box containing the maser emission was chosen to obtain the spectral data presented in Fig.\ref{spec}. We use the Continuum and Line Analysis Single-dish Software (\textsc{Class}), which is part of the \textsc{Gildas} software package, to fit the spectrum of each maser spot and get its peak flux density, peak velocity, line width and integrated flux density with Gaussians.

In the case of 1.1-3.1\,GHz broad-band (continuum) data, we also used the \miriad package to reduce the data. Dirty maps of the continuum data were obtained with the task \textsc{invert}, applying frequency synthesis and a robust parameter of 0.5. Beam sizes at the two epochs were $7.3\arcsec\times3.9\arcsec$ and $12.1\arcsec\times4.6\arcsec$, respectively. Images were subsequently deconvolved with the task \textsc{mfclean} to perform the CLEAN algorithm. Continuum data at distances $<$ 3\,k$\lambda$ in the uv-plane were discarded in the second epoch to filter out extended Galactic emission. 

\section{Results}
\label{result}

\subsection{Continuum emission}
\label{continuum}

The radio continuum emission at 18\,cm is unresolved at both epochs. The peak of the radio continuum emission is located at RA(J2000) = 16$\rm ^{h}$37$\rm ^m$06$\rm ^s$.60, DEC(J2000) = $-$48$^\circ$13$^{\prime}$42$^{\prime\prime}$.6 with a mean flux density between 1.1-3.1\,GHz of 17.11$\pm$0.09\,mJy at the first epoch and RA(J2000) = 16$\rm ^{h}$37$\rm ^m$06$\rm ^s$.60, DEC(J2000) = $-$48$^\circ$13$^{\prime}$42$^{\prime\prime}$.7 with a mean flux density between 1.1-3.1\,GHz of 20.78$\pm$0.22\,mJy at the second epoch. The difference in flux density between two epochs may be caused by the different array configurations with the 1.5C configuration tracing more extended structures. The errors given in the paper are all at a 1-$\sigma$ level.

\begin{figure}
\begin{center}
\includegraphics[width=0.45\textwidth]{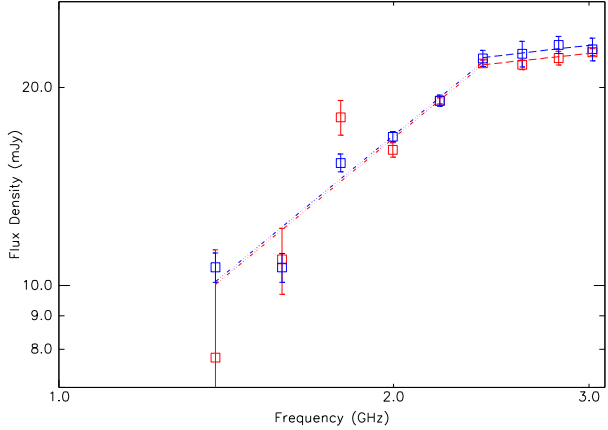}
\caption{Variation of flux density of radio continuum emission as a function of frequency. Red symbols and blue symbols are from the first epoch and second epoch, respectively. The dot-dashed lines are the fitted results from 1383\,MHz to 2407\,MHz with spectral indices of 1.39$\pm$0.11 (first epoch) and 1.39$\pm$0.08 (second epoch), respectively. The dashed lines are the fitted results from 2407\,MHz to 3022\,MHz with spectral indices of 0.18$\pm$0.07 (first epoch) and 0.20$\pm$0.19 (second epoch), respectively. Error bars are 1-$\sigma$ errors.}
\label{spein}
\end{center}
\end{figure}

The flux density of the source shows significant variation across the observed 1.1-3.1\,GHz band. Fig.\ref{spein} shows the flux density in 0.2-GHz intervals across this band. At both epochs the flux density increases with frequency, but this trend is steeper for frequencies $\nu \la 2.4$\,GHz. The derived spectral indices $\alpha$ (with $S_\nu\propto \nu^\alpha$, where $S_\nu$ is the flux density) are 1.39$\pm$0.11 (first epoch) and 1.39$\pm$0.08 (second epoch) between $\simeq 1.4$ and 2.4\,GHz, and 0.18$\pm$0.07 (first epoch) and 0.20$\pm$0.19 (second epoch) between $\simeq 2.4$ and 3.0\,GHz. The derived spectral indices are consistent, within errors, for both epochs. Since these observations were obtained with different array configurations, we believe the flux density trend is a robust result and not the consequence of extended emission being missed at the higher frequencies. We discuss the spectral index of the source in more detail in Section \ref{nature}. 

The spectrum of the radio continuum emission in IRAS 16333$-$4807 is consistent with free-free emission in an ionized region. At low frequencies, a spectral index $\alpha\simeq 1.39$ indicates partially optically thick free-free emission. Above the turnover frequency of $\simeq 2.4$\,GHz, the emission is optically thinner, with a spectral index $\alpha\simeq 0.2$, which is close to the optically thin limit in an ionized region ($\simeq -0.1$). We note, however, that the flux density at 22\,GHz reported by \citet{Use2014} is not consistent with the spectrum in our observation. A crude extrapolation with $\alpha \simeq 0.2$ to 22\,GHz would give a flux density of $\simeq 27$ mJy. Therefore, there is an excess of $\simeq 70$ mJy in the \citet{Use2014} observations. This discrepancy may be due to source variability. Note that we do not expect the spectral index of free-free emission to be $\simeq 0.2$ all the way to 22 GHz. At high frequencies this emission will become optically thin, and it will flatten up to $\alpha\simeq -0.1$. In that case, the excess of emission at 22 GHz will be even higher. Alternatively, it is possible that the emission at 22\,GHz includes a contribution additional to free-free radiation from the photoionized nebula, such as a central ionized disk \citep{Tae2009}, a collimated wind \citep{Ree1986}, or circumstellar dust \citep{Cee2008}. A more complete frequency coverage above 3\,GHz is necessary to better ascertain the nature of the radio continuum emission.

\subsection{Maser emission}
\label{maser}

\begin{figure}
\begin{center}
\includegraphics[width=0.45\textwidth]{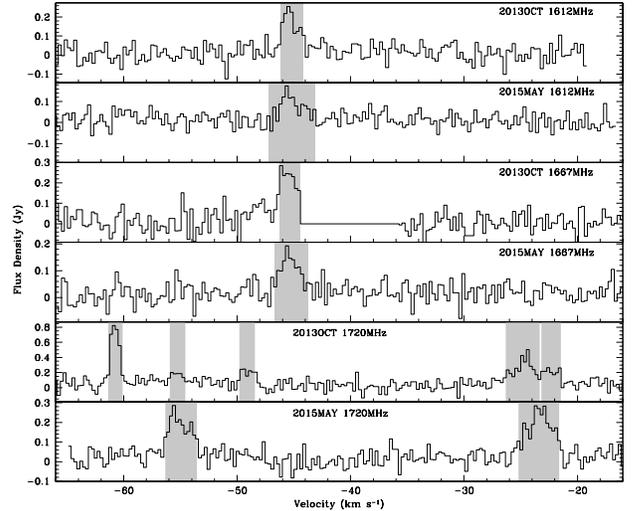}
\caption{OH maser spectra of IRAS 16333$-$4807 at different transitions and epochs. The shaded velocity range shows the emission channels for each maser spot (spectral component). The blank channels (from about $-$45\,\kms to $-$35\,\kms) in the 1667\,MHz spectrum taken in October 2013 were due to RFI and prompted us to re-observe this source.}
\label{spec}
\end{center}
\end{figure}

Fig.\ref{spec} shows the OH maser spectra of this source at two epochs, smoothed to a  spectral resolution of $\simeq$0.27\kms. We detect no emission from the 1665\,MHz transition, but do find emission in the 1612, 1667 and 1720\,MHz transitions at both epochs. The shaded velocity range shows the emission channels for each maser spot\footnote{the spectral peak considered to arise in a single, well-defined position \citep{Wae2014}}. The 1720\,MHz OH maser shows variability during the two epochs. In the single-dish Parkes observations from the SPLASH survey \citep{Dae2014}, the data show wide absorption and emission features and we were unable to use them to obtain additional information on variability of the OH maser emission. These detections are very unusual for a PN. In particular, they confirm that IRAS 16333$-$4807 is the seventh OHPN, the second OHPN with 1720\,MHz OH maser emission, and \textit{the only evolved object in whose circumstellar envelope the 1720\,MHz maser is the strongest OH transition}.

\begin{table*}
\begin{center}
\caption{Parameters of OH maser spots in IRAS 16333$-$4807.}
\label{maintable}
\begin{tabular}{ccccccccccc}
\hline
\hline
RA&DEC&Frequency&$\sigma_{\rm min}$&$\sigma_{\rm max}$&Position&$S_\nu$&$V_{\rm LSR}$&Line&Integrated\\
(J2000)&(J2000)&&&&Angle&&&Width&$S_\nu$\\
($^{\rm h\,m\,s}$)&($^{\circ~\prime~\prime\prime}$)&(MHz)&(arcsec)&(arcsec)&($^{\circ}$)&(Jy)&(\kms)&(\kms)&(Jy\kms)\\
\hline
2013OCT&&&&&&&&&\\
16:37:06.57&-48:13:42.4&1612&0.3&0.5&-9.6&0.19&-45.2$\pm$0.1&2.1$\pm$0.4&0.43$\pm$0.06\\
16:37:06.57&-48:13:42.3&1667&0.4&0.6&-4.6&0.32&-45.5$\pm$0.1&1.8$\pm$0.2&0.60$\pm$0.06\\
16:37:06.59&-48:13:42.4&1720&0.3&0.5&-4.6&0.79&-60.73$\pm$0.02&0.85$\pm$0.05&0.71$\pm$0.04\\
16:37:06.62&-48:13:42.6&1720&0.6&0.9&-4.6&0.13&-55.3$\pm$0.2&1.1$\pm$0.3&0.16$\pm$0.04\\
16:37:06.54&-48:13:42.4&1720&0.5&0.7&-4.6&0.16&-49.0$\pm$0.1&1.2$\pm$0.2&0.20$\pm$0.04\\
16:37:06.63&-48:13:42.5&1720&0.3&0.6&-4.6&0.27&-24.7$\pm$0.1&1.7$\pm$0.2&0.48$\pm$0.05\\
16:37:06.65&-48:13:41.5&1720&0.4&0.6&-4.6&0.16&-22.9$\pm$0.2&2.0$\pm$0.4&0.33$\pm$0.06\\
2015MAY&&&&&&&&&\\
16:37:06.62&-48:13:43.3&1612&1.6&2.9&-21.1&0.12&-45.2$\pm$0.2&2.1$\pm$0.4&0.27$\pm$0.04\\
16:37:06.49&-48:13:42.3&1667&1.3&2.3&-21.9&0.14&-45.4$\pm$0.1&2.0$\pm$0.3&0.31$\pm$0.04\\
16:37:06.59&-48:13:43.0&1720&1.0&1.9&-21.9&0.26&-55.1$\pm$0.1&2.4$\pm$0.3&0.66$\pm$0.06\\
16:37:06.66&-48:13:42.9&1720&0.9&1.5&-21.9&0.29&-23.2$\pm$0.1&2.5$\pm$0.2&0.78$\pm$0.06\\

\hline
\end{tabular}

Notes: Columns 1 and 2 are fitted positions (RA and DEC). Column 3 is the frequency. Columns 4, 5 and 6 are minor axis uncertainty, major axis uncertainty and position angle (measured from north to east). Columns 7 and 8 are peak flux density and peak velocity. Column 9 is line width. Column 10 is integrated flux density. The positions and relative positional uncertainty are derived from the \miriad using \textsc{imfit} task. The peak flux density, peak velocity, line width and integrated flux density are derived from Gaussian fitting in \textsc{Class}.

\end{center}
\end{table*}

Table \ref{maintable} shows the relevant parameters of each maser spot at both epochs. The emission at 1612 and 1667\,MHz shows a single spectral component at both epochs, peaked at $\simeq -45$\kms~in all cases. This velocity is close to that of the \water maser emission \citep[two components at $\simeq -41.8$ and $-43.9$\kms]{Use2014} and it is likely to be close to the stellar velocity. However, the 1720\,MHz spectra show spectral components significantly offset, both red- and blue-shifted from that velocity. We find five maser spots in the October 2013 epoch, with velocities of approximately $-$60.7, $-$55.3, $-$49.0, $-$24.7 and $-$22.9\,\kms. During the May 2015 observation, we only find two maser spots at velocities of $\sim-$55.1 and $-$23.2\,\kms.

\begin{figure}
\begin{center}
\includegraphics[width=0.45\textwidth]{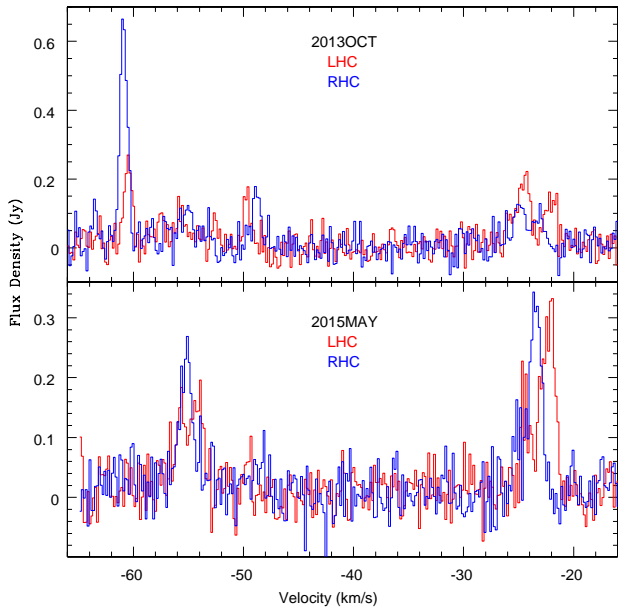}
\caption{The LHC and RHC polarization spectra of the 1720\,MHz transition at two epochs. The velocity resolution is 0.18\kms.}
\label{LR}
\end{center}
\end{figure}

Fig.\ref{LR} shows the right- and left-hand circular (RHC, LHC) polarization spectra of the 1720\,MHz transition, where we adopt the IEEE convention for polarization handedness. There is a clear frequency shift between the RHC and LHC polarizations in all components, indicating Zeeman splitting due to a significant magnetic field. The derived magnetic field along the line of sight, assuming a Lande G factor of $\simeq 0.113$ km s$^{-1}$ mG$^{-1}$ \citep{PZ1967,Coe1975}, is given in Table \ref{magtable}. Note that the RHC polarization is at a higher frequency than the LHC for most components, indicating that the magnetic field is oriented toward us \citep{Gre2014}. The only exception is the maser spot at $\simeq -49.0$\kms, whose shift indicates that the magnetic field is pointing away from us. We do not detect any significant Zeeman splitting in the 1612 and 1667\,MHz transitions. The magnetic field upper limits for the 1612 and 1667\,MHz transitions are 2.1\,mG and 0.6\,mG, respectively.

\begin{table*}
\begin{center}
\caption{Magnetic field properties of the 1720\,MHz OH maser components in IRAS 16333$-$4807.}
\label{magtable}
\begin{tabular}{ccccccc}
\hline
\hline
$V_{\rm LSR}$&Pol&$S_{\rm p}$&&Magnetic field&\\
&&&Number&Strength&Reliability\\
(\kms)&&(Jy)&&(mG)&\\
\hline
2013OCT&&&&\\
-60.40&LHC&0.23&$Z_{1}$&$-3.6\pm0.9$&A\\
-60.81&RHC&0.64&$Z_{1}$&&\\
-49.61&LHC&0.16&$Z_{2}$&$+7.8\pm1.1$&A\\
-48.73&RHC&0.16&$Z_{2}$&&\\
-24.51&LHC&0.17&$Z_{3}$&$-2.7\pm1.3$&A\\
-24.81&RHC&0.11&$Z_{3}$&&\\
-22.00&LHC&0.13&$Z_{4}$&$-11.8\pm2.2$&C\\
-23.34&RHC&0.07&$Z_{4}$&&\\
2015MAY&&&&\\
-54.83&LHC&0.14&$Z_{1}$&$-2.5\pm1.4$&C\\
-55.11&RHC&0.21&$Z_{1}$&&\\
-22.29&LHC&0.28&$Z_{2}$&$-9.7\pm1.1$&A\\
-23.39&RHC&0.32&$Z_{2}$&&\\
\hline
\end{tabular}

Notes: Column 1 is the local standard of rest (LSR) velocity of the component peak, which is derived from least-squares fitted Gaussian components with \textsc{Class}. Column 2 and 3 are the polarisation and peak flux density of the component. Column 4 is the numbered Zeeman pairs. Column 5 is the magnetic field strength with errors, which are all at 1-$\sigma$ level. We adopt the common field convention with RHC at a lower velocity than LHC representing a negative magnetic field strength, directed towards us. The last column is the reliability for the Zeeman pairs, which is defined according to \citet{Gre2015}, with class `A' being the most reliable. Note that we did not obtain the magnetic field strength of the weak maser component at the velocity of $-$55.3\kms~from the first epoch, as the signal to noise ratio for this maser is low, and as such the splitting has low statistical significance.

\end{center}
\end{table*}

\begin{figure}
\begin{center}
\includegraphics[width=0.45\textwidth]{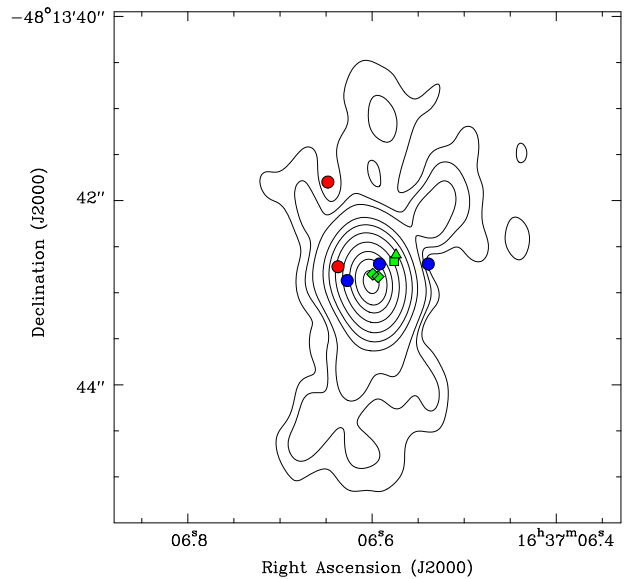}
\caption{Distribution of the maser spots with respect to the lobes of the PN. Contours represent the radio continuum emission of IRAS 16333$-$4807 at 1.3\,cm from \citet{Use2014}. The \water maser components are represented with green filled diamonds. The 1612 and 1667\,MHz OH maser spots from this paper are represented with green filled triangle and green filled square. The blue and red filled circles are blue-shifted and red-shifted 1720\,MHz OH maser spots compared to the velocities of \water, 1612 and 1667\,MHz maser spots. Note that only the OH maser spots from the first epoch are shown in this figure because they have higher absolute positional accuracy than those from the second epoch. The positional uncertainties of maser spots are not plotted but are listed in Table \ref{maintable}.}
\label{contour}
\end{center}
\end{figure}

Fig.\ref{contour} shows the locations of the OH maser emission with respect to the lobes of the nebula, traced by a radio continuum map at 1.3\,cm \citep{Use2014} .
The 22\,GHz \water maser spots from \citet{Use2014} are also plotted. We only show the OH maser spots from the first epoch in this paper, as these have higher absolute positional accuracy than those of the second epoch. To obtain a more accurate comparison between our data and those in \citet{Use2014}, we assumed that the absolute position of the radio continuum emission is the same at both 18\,cm and 1.3\,cm, and shifted our 18 cm data (both continuum and OH maser spots) accordingly. The two continuum peaks are within the 1\arcsec~absolute positional uncertainty of the ATCA observations. Most maser spots are close to the center of the PN and they do not seem to be aligned along the lobes, but rather trace a structure perpendicular to them. The only exception could be the 1720\,MHz maser spot at $\simeq -$22.9\kms, which seems to be located towards the northern lobe of the PN.

\section{Discussion}
\label{discussion}

\subsection{The evolutionary stage of IRAS 16333$-$4807}
\label{nature} 

The presence of maser emission (OH or H$_2$O) in a PN is usually considered to be a sign of its youth, since these masers are expected to disappear shortly after the end of the AGB phase \citep{Goe1990}. \citet{Use2012} find the sizes and optical visibilities of the maser-associated PNe suggest a range of ages, but it is possible that the subgroup of PNe showing both OH and H$_2$O are among the youngest PNe known \citep{Goe2015}. Our observations confirm IRAS 16333$-$4807 as a member of this rare group of potentially young PNe. So far, in addition to IRAS 16333$-$4807, only K 3$-$35 \citep{Mie2001}, IRAS 17347$-$3139 \citep{Deg2004,Tae2009}, and possibly IRAS 17347$-$3139 \citep{Goe2015} show maser emission from both molecules.

The spectral energy distribution of the radio continuum emission can also provide some information about the evolutionary stage of the source. As mentioned in section \ref{continuum}, the radio continuum spectrum is compatible with free-free radiation, and it is similar to that seen in other PNe \citep{Aak1991}. There seems to be a turnover frequency of $\simeq 2.4$\,GHz, where the opacity regime changes from optically thick to optically thin emission. The turnover frequency is directly related to the emission measure of the ionized region. In this case, we derive an emission measure of 1.9$\times10^{7}$ pc\,cm$^{-6}$ obtained from the free-free opacity $\tau_{\nu}\approx3.28\times10^{-7}(\frac{\rm T_{e}}{10^{4}{\rm K}})^{-1.35} (\frac{\nu}{\rm GHz})^{-2.1}(\frac{\rm EM}{\rm pc\,cm^{-6}})$, with $\tau_\nu\simeq 1$ at $\nu=2.4$\,GHz and assuming a typical equilibrium temperature of T$_{\rm e}\sim10^{4}$ K. Compared to the other PNe with OH and \water masers, the turnover frequency in IRAS 16333$-$4807 is lower. This frequency is $\simeq$10\,GHz for K 3$-$35 \citep{Aak1991}, and $\ga 20$\,GHz for IRAS 17347$-$3139 \citep{Goe2005,Tae2009}. This means that the emission measure of the ionized gas seems to be lower in IRAS 16333$-$4807. A high emission measure could be a signature of youth \citep{Kwo1981}. Therefore, assuming that the central stars in the group of PNe with both OH and H$_2$O masers have similar characteristics, the turnover frequency suggests that IRAS 16333$-$4807 is somewhat more evolved than K 3$-$35 and IRAS 17347$-$3139.

\subsection{The origin of OH maser emission}
\label{origin}

The most striking characteristic of IRAS 16333$-$4807 is the presence of OH maser emission at 1720\,MHz, and that this emission is stronger than the other ground-state OH transitions.
It was proposed by \citet{El1976} that strong 1720\,MHz maser emission can only be produced by means of collisional pumping under particular physical conditions with T$_k$ $\leq$ 200 K and n$_{\rm H_{2}} \simeq$ 10$^{5}$ cm$^{-3}$.
It is likely a shock is required to produce 1720\,MHz OH masers, as has been found in the case of SNRs. However, strong (J-type) shocks tend to break apart molecules, which then reform behind the shock front, but do not produce sufficient column for masers to appear during the rapid cooling phase \citep[e.g.][]{Loe1999,War2012}. Thus, C-type shocks are more likely to be conducive to maser conditions. \citet{Dre1980} showed that in the presence of strong magnetic fields, shocks would be of the non-dissociative C-type, due to the interaction of the ions upstream from the shock front with the magnetic field. This appears to be a valid mechanism for producing such shocks in IRAS 16333$-$4807, since we do measure significant magnetic fields, of order a few milliGauss.
The other ground-state OH maser transitions are excited radiatively. IRAS 16333$-$4807 also harbours \water masers, which can be excited in C-shocks \citep{Hoe2013}. Our measurement of Zeeman splitting in the 1720\,MHz transition is consistent with this maser being pumped in a magnetized environment. The magnetic field strength of the $-$60.7\kms~component at the first epoch is comparable to that of the $-$55.1\kms~component at the second epoch within the errors (Table \ref{magtable}), which suggests that they are tracing the same structure.

It is believed that PNe are shaped by jets ejected during the post-AGB phase, which carve cavities in the circumstellar envelope \citep{Sah1998}. These jets obviously create shocks, which can power maser emission. This is clearly seen in the case of the post-AGB stars called ``water fountains'', in which \water masers trace collimated jets \citep{ima07}.  If the OH masers at 1720\,MHz in IRAS 16333$-$4807 were pumped by jets, we would expect them to be associated with the lobes of the PN. This could be the case for the 1720\,MHz maser at $\simeq -$22.9\kms. However, most of these masers, and the overall distribution of all OH and H$_2$O masers (Fig.\ref{contour}) seem to trace an equatorial structure, perpendicular to the lobes. This could be related to the low-velocity equatorial flows seen in some ``water fountains'' \citep{ima07} and the toroidal distribution of \water masers in other maser-emitting PNe \citep{Deg2004,Use2008}. The spatial distribution of the 1720\,MHz OH maser spots and the presence of clear Zeeman pairs suggest that OH emission at 1720\,MHz in IRAS 16333$-$4807 arises from an energetic ejection of material along the equatorial plane, probably in the form of an expanding, magnetized torus. Note that K 3$-$35, the other PN showing maser emission at 1720\,MHz, has also been inferred to host  a magnetized disk \citep{Mie2001,Goe2009}. These equatorial ejections could be created, for instance, by binary interactions \citep{Noe2006,Hue2007}. 
Magnetized equatorial structures traced by maser emission have also been found in post-AGB stars \citep{Bae2003,Bae2004}, with similar values for the derived magnetic field as the ones found in PNe \citep[][and this paper]{Goe2009}. This suggests that the characteristics of the magnetic field do not significantly evolve in the transition to the PN phase \citep{Vlv2008}.

The presence of OH emission at 1720\,MHz could be the hallmark of a particular phase of PN evolution, characterized by energetic equatorial ejections. The extreme rarity of this emission in evolved stars suggests that this phase is very short, or that energetic equatorial ejections only happen in the form of recurrent, but short-lived events during the early PN phase. 

It is interesting to note the different velocity pattern and magnetic field properties of the OH maser emission at 1720\,MHz, compared with the other OH transitions (Fig.\ref{spec}, \ref{LR}) and the H$_2$O emission \citep{Use2014}. The emission of all these other transitions clusters around $-45$\kms , while OH emission at 1720\,MHz shows components both red- and blue-shifted with respect to that velocity. If the 1720\,MHz masers are tracing an expanding torus, its expansion velocity would be at least $\simeq 20$\kms~(half of the maximum velocity spread of the maser components). The OH masers at 1612 and 1667\,MHz, as well as the H$_2$O masers seem to be closer to the star (Fig.\ref{contour}), but these masers have lower relative velocities than those at 1720\,MHz. If the maser transitions arise from the same equatorial structure, this velocity pattern may indicate that the expansion velocities are higher at the outer parts of the torus, which would be expected if the ejection had an explosive nature. Alternatively, it could be a projection effect, if the OH masers at 1720\,MHz tend to trace the expansion along the line of sight, and the other maser transitions are seen in regions expanding along the plane of the sky. This could happen if the velocity gradient is very high in the central regions, which would not favour the maser amplification along the line of sight. Moreover, the magnetic field is stronger in the region emitting at 1720\,MHz than for the other OH masers, which suggests that these transitions are pumped in regions with significantly different magnetic properties. This suggests that the different velocity patterns in these OH transitions are not merely projection effects, but that the 1720\,MHz masers are selectively tracing the energetic ejection of an equatorial magnetized structure. Further observations should be undertaken to obtain a better understanding of the source.

\subsection{Comparison with K 3$-$35}
\label{comparison} 

K 3$-$35 was the only known PN showing 1720\,MHz OH masers before our work on IRAS 16333$-$4807. Since K 3$-$35 and IRAS 16333$-$4807 are the only PNe showing OH emission at 1720\,MHz, we carry out a comparison between the two sources, to understand whether they can define a special group of sources in terms of common physical characteristics. The ground-state OH maser emission, however, show significant differences between both sources: 

(1) K 3$-$35 has multiple spectral components in the 1612 and 1667\,MHz transitions and only one spectral component at 1720\,MHz, while IRAS 16333$-$4807 shows the opposite behaviour with only one spectral component at 1612 and 1667\,MHz and multiple spectral components at 1720\,MHz. 

(2) In K 3$-$35, the spatial distribution of the masers at 1720\,MHz is more compact than the 1612 and 1667\,MHz masers, while in IRAS 16333$-$4807, the 1720\,MHz OH maser components are more extended.

(3) The velocity structure of the masers in these sources is more difficult to determine, since the stellar velocity is highly uncertain in both cases. In IRAS 16333$-$4807, we assume a stellar velocity of $\simeq -45$\kms (Section \ref{maser}). Under this assumption, OH masers at 1612 and 1667\,MHz would be close to the stellar velocity, and there are both red- and blue-shifted components of emission at 1720\,MHz. In the case of K 3$-$35, we favour a stellar velocity of $\simeq +10$\kms, based on optical spectroscopy \citep{Mie2000}, rather than the +23\kms quoted by other authors \citep[e.g.][]{Goe2009}, based on the presence of a broad CO component at that velocity \citep{Tae2009}. Interferometric CO observations \citep{SS2012} show that the gas at +23\kms is located west of the central star, while the emission centered at K 3$-$35 is between +9 and +18\kms. Under this assumption (stellar velocity of +10\kms), the emission at 1667\,MHz  and the strongest component at 1612\,MHz are close to the stellar velocity, while 1665 and 1720\,MHz emission is red-shifted. Additional components of OH at 1612\,MHz are both red- and blue-shifted. All this indicates a complicated kinematic pattern, but no clear similarity between K 3$-$35 and IRAS 16333$-$4807 arises. 

These spectral, spatial and kinematic differences may be caused by different mass-loss processes in these two PNe. Alternatively, it is possible that they reflect a time variation in intrinsically similar processes, such as recurrent episodes of mass loss. Monitoring of the OH emission in these sources may help to clarify this issue.

\section{Conclusions}
\label{conclusion}

OH maser emission at 1612, 1667 and 1720\,MHz have been detected toward the PN IRAS 16333$-$4807. The pattern of this maser emission is very unusual in a PN: it is the seventh confirmed PN with OH maser emission, the second one exhibiting 1720\,MHz maser emission and the only evolved stellar source with 1720\,MHz OH maser as the strongest transition. The 1720\,MHz transition has a wide velocity range compared to the 1612 and 1667\,MHz transitions and shows variability at two epochs. The presence of both OH and H$_2$O maser emission suggests that it is one of the youngest PNe known, although it may be slightly more evolved than other sources showing these masers. We detect Zeeman splitting in the 1720\,MHz transition at two epochs, which suggests that the 1720\,MHz OH masers are formed in a magnetized environment. These 1720\,MHz OH masers in PN IRAS 16333$-$4807 may trace short-lived equatorial ejections around the formation time of the PN.

\acknowledgments The Australia Telescope Compact Array is part of the Australia Telescope which is funded by the Commonwealth of Australia for operation as a National Facility managed by CSIRO. This research has made use of NASA’s Astrophysics Data System Abstract Service. We thank Dr. Lucero Uscanga for allowing us the use of the image at 1.3 cm. HHQ would thank the China Scholarship Council (CSC) support. JFG is partially supported by MICINN (Spain) grants AYA2011-30228-C03-01 and AYA2014-57369-C3-3 (co-funded by FEDER).

\end{document}